\documentclass[12pt]{article}
\usepackage{epsfig}
\usepackage{graphicx}

\def\beq{\begin{equation}}
\def\eeq#1{\label{#1}\end{equation}}
\def\eeqn{\end{equation}}

\def\beqa{\begin{eqnarray}}
\def\eeqa#1{\label{#1}\end{eqnarray}}
\def\eeqan{\end{eqnarray}}

\let\bar=\overbar

\def\Dslash{\not{\hbox{\kern-4pt $D$}}}
\def\dslash{\not{\hbox{\kern-2pt $\del$}}}

\def\msb{{\bar{\ssstyle M \kern -1pt S}}}

\usepackage{fancyhdr,graphicx}
\def\Title#1{\begin{center} {\Large {\bf #1} } \end{center}}

\begin{document}
\Title{Condition for the burning of hadronic stars into quark stars}

\bigskip\bigskip


\begin{raggedright}
{\it A. Drago and G. Pagliara}
\end{raggedright}

\begin{raggedright}
{\it Dip.~di Fisica e Scienze della Terra dell'Universit\`a di Ferrara and INFN
Sez.~di
Ferrara, Via Saragat 1, I-44100 Ferrara, Italy}
\end{raggedright}

\section{Introduction}
The question about the existence of absolutely stable strange quark
matter in astrophysical compact objects is still debated in a number
of theoretical and observational investigations, see
Ref. \cite{Buballa:2014jta} for a recent review. In particular the
formation of quark stars is a very interesting issue.  In the seminal
paper \cite{Olinto:1986je}, the process of conversion of a neutron
star into a quark star was shown to be a very exothermic process
which, if it really occurs in Nature, should lead to powerful
explosive phenomena, maybe also connected with supernovae and
gamma-ray-bursts
\cite{Berezhiani:2002ks,Drago:2004vu,Drago:2008tb}. The relevance of
this process, from the phenomenological point of view, is related to
the time scale of the conversion. Only if sufficiently fast (with time
scales of the order of tens of seconds or less) the formation of a quark star
could release detectable signals; on the other hand, a very slow
conversion would not probably provide any evident signature. Interestingly, semi-analytical estimates
and numerical hydrodynamics simulations have provided hints for a very
fast conversion, occurring on time scales of the order of ms
\cite{Lugones:1994xg,Drago:2005yj,Niebergal:2010ds,Herzog:2011sn,Pagliara:2013tza}.
In this short contribution, we will review the approach used for
studying the conversion of a neutron star into a quark star based on
the assumption of a infinitely thin combustion zone and we will
discuss why, in this scheme, the combustion stops before the whole
hadronic star is converted.

\section{Combustion of hadronic stars}
The process of combustion is a very complicated phenomenon which in
principle must be modeled by coupling the hydrodynamic equations for the
mixture of the two fluids 
(the fuel and the ashes) with the equations of conservation
of chemical species in which both diffusion processes and rates of
chemical reactions are included \cite{williams}, see \cite{Horvath:2007tv,Niebergal:2010ds} for the case of quark stars.  
A common assumption, especially used in astrophysical problems such as type Ia supernovae \cite{Reinecke:1998mk}, consists in treating the
combustion zone as a surface of discontinuity separating the fuel from
the ashes. This assumption is necessary since
in many cases
the width of the combustion zone is much smaller then the size of the system
and it would be computationally unfeasible to resolve the microscopic dynamics 
of the combustion zone. 
As in the case of shock waves, in presence of a discontinuity,
hydrodynamic equations lead to the conditions 
of continuity of the
baryon flux, the momentum flux and the energy flux across the burning
front \cite{landau}.  
Indicating with $e_i$, $p_i$, $n_i$, $w_i$,
$X_i=(e_i+p_i)/n_i^2$ the energy density, pressure, baryon density,
enthalpy and dynamical volume of the i-th fluid, those conservation
laws, generalized to relativistic hydrodynamics \cite{coll}, read:
\begin{eqnarray}
n_1 u_1&=&n_2 u_2=j\\
(p_2-p_1)/(X_1-X_2)&=&j^2\\
w_1(p_1,X_1)X_1-w_2(p_2,X_2)X_2&=&(p_1-p_2)(X_1+X_2)
\end{eqnarray}
where $u_i$ are the four-velocities of the two fluids in the front rest
frame and $j$ is the baryon number flux across the front and it 
must be determined within a microscopic kinetic
approach, as the one of Ref. \cite{Olinto:1986je}. 
This system of equations allows
to determine the velocities of the two fluids and the state of fluid 2
once the initial state of fluid 1 is fixed. Depending on the values
of the velocities $u_i$ and the sound velocities of the fluids $c_i$,
one can obtain detonations, which are processes of
combustion driven by a shock wave, or deflagrations in which instead
combustion proceeds thanks to the diffusion of heat or of chemical
species \cite{landau}.

The condition of exothermic combustion, generalized to relativistic hydrodynamics by Coll in 
Ref. \cite{coll}, reads $\Delta(p,X)=e_1(p,X)-e_2(p,X)=w_1(p,X)-w_2(p,X)>0$ and
allows to find the window of baryon densities of fluid 1 for which the
combustion can proceed.  We want here to clarify the meaning of this
condition.  Let us fix the initial state A of fluid 1: $p_1=p_A,
e_1=e_A, X_1=X_A$ (the temperature of fluid 1 is set to zero).  From this point one can draw the shock adiabat of
fluid 1 in the (p,X) plane.  Moreover, by using the equation of state
of fluid 2 and Eq.3 one can draw in the same plane the detonation
adiabat, see Fig.1.
In the following we will demonstrate that, the Coll's condition implies that the shock adiabat lies 
below the detonation adiabat. 
Let us assume that the equation of state of fluid 2 is a generic 
polytrope
$e_2=\alpha n_2 + p_2/(\gamma-1)$ (\cite{Ozel:2010fw}), $p_2=k n_2^\gamma$, where $\gamma$ is the adiabatic index
and $1<\gamma \leq 2$ (the second inequality implying that the equation of state is causal at all densities \cite{Read:2008iy}). 

One can derive the following expression for the energy 
density as a function of $p_2$ and $X_2$:
\begin{equation}
e_2(p_2,X_2)=\frac{\alpha^2(\gamma-1)+2p_2X_2+\alpha\sqrt{\gamma-1}\sqrt{\alpha^2(\gamma-1)+4\gamma p_2X_2}}{2X_2(\gamma-1)}
\end{equation}

\begin{figure}[ptb]
\vskip 1cm
\begin{centering}
\epsfig{file=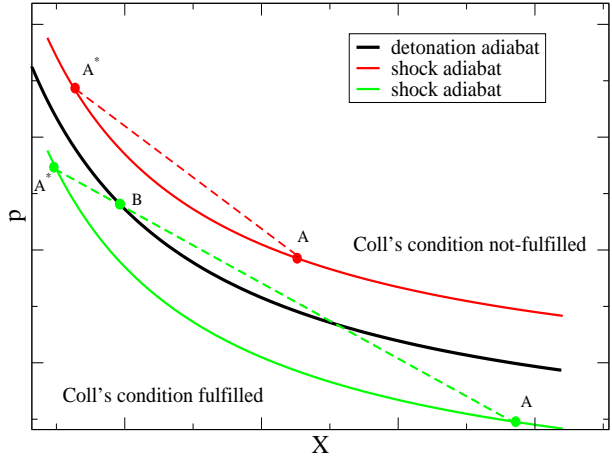,height=8cm,width=6cm,angle=-90}
\caption{Illustrative plot of the shock and the detonation adiabats 
in the case in which the Coll's condition is fulfilled or not.}
\label{soglie}
\end{centering}
\end{figure}

Let us fix $X_1=X_2=X_A$ and assume $\Delta(p_A,X_A)>0$.
If $p_2 > p_1=p_A$, the detonation adiabat lies above the shock adiabat.
With this setting, the detonation adiabat reads (adding and subtracting $e_2(p_A,X_A)$):
\begin{equation}
\Delta(p_A,X_A)=p_A-p_2+e_2(p_2,X_A)-e_2(p_A,X_A)
\end{equation}
which after some manipulation and using Eq.4 reads:
\begin{eqnarray}
\Delta(p_A,X_A)&=&\frac{\alpha}{2X_A\sqrt{\gamma-1}}\left(\sqrt{\alpha^2(\gamma-1)+4\gamma p_2 X_A}
-\sqrt{\alpha^2(\gamma-1)+4\gamma p_A X_A}\right)\\&+&(p_2-p_A)\frac{2-\gamma}{\gamma-1}
\end{eqnarray}
Since $1<\gamma \leq 2$ the sign of $\Delta(p_A,X_A)$ clearly determines the sign of $p_2-p_A$.
Thus, if $\Delta(p_A,X_A)>0$, i.e. if the Coll's condition holds true,
the initial point A lies below the detonation adiabat \footnote{For values of $\gamma>2$, one can still obtain the same 
conclusion but depending on the specific values of $\alpha$, $p_A$ and $X_A$}.
Since the detonation adiabats and the shock adiabats do not cross in the (p,X) plane \cite{landau}
this implies that the whole shock adiabat lies below the detonation adiabat (at least in the standard cases).
 The Coll's condition is necessary for obtaining detonations. Detonation is a process of
combustion which is driven by a shock wave propagating within the
fuel: when the shock wave passes through A, the fluid is compressed
and heated up to the state A$^*$ which lies on the shock adiabat (see Fig.1). The
chemical reactions start, the fluid expands and cools down until the
combustion is complete and the state B, lying on the detonation
adiabat, is reached \cite{landau}.  On the other hand, if the shock
adiabat of the fuel lies above the detonation adiabat, it is not
possible to trigger the combustion via a shock wave. A shock wave
passing through the point A would heat up the matter but not at a
sufficiently high temperature to start the combustion (see Fig.1).
Once the Coll's condition is fulfilled the hydrodynamical combustion
can proceed either as a detonation or a deflagration depending on the
specific microphysics and on the equation of state.
The equation $\Delta(p_A,X_A)=0$ allows to find the state A, 
characterized by its baryon density  $n_1^*$, below which 
the combustion cannot proceed anymore. At this density 
$e_1=e_2=e_A$, $p_1=p_2=p_A$ which together with $X_1=X_2=X_A$ also implies 
$n_1=n_2=n_1^*$. At this value of baryon density there is no surface of discontinuity anymore
and the two phases are in mechanical equilibrium.
Previous studies 
have shown that, for many equations of state, $n_1^* \sim 0.2-0.3$ fm$^{-3}$ \cite{Lugones:1994xg,Herzog:2011sn}
and indeed the fast combustions found in the numerical simulations of Ref. \cite{Herzog:2011sn}
stop exactly at those values of density.

One has to remind that the scheme based on a infinitely thin combustion zone 
is only an approximation. Relieving this approximation and considering 
the microscopic processes of diffusion of quarks and reactions
between quarks (such as $u+d->u+s$) occurring in the finite width combustion
zone allows to follow the subsequent evolution of the system which proceeds 
until the whole star is converted \cite{last}.

\section{Conclusions}
We have discussed the approximation scheme of combustion based on the
assumption of a infinitely thin combustion layer. This model has been
widely used in numerical simulations of type Ia supernovae
\cite{Reinecke:1998mk} and recently also for numerical investigations of the conversion of
hadronic stars into quark stars \cite{Herzog:2011sn,Pagliara:2013tza}.
At densities larger than the critical density $n_1^*$, for which the
Coll's condition is fulfilled, the process of conversion proceeds very
fast with the effective velocity of conversion significantly augmented
by Rayleigh-Taylor instabilities. After a few ms, when the conversion
front reaches $n_1^*$, a big part of the star is converted but a few
$0.1 M_{\odot}$ remain unburnt and will convert on a longer time
scale, of the order of tens of seconds \cite{last}.  Note that, up to
now, the conversion process has been studied only within cold and non
rotating hadronic stars. The dynamics of the birth of quark stars can
be qualitatively different in other astrophysical situations such as
supernovae and protoneutron stars and mergers of neutron stars. Those
cases are, to date, essentially unexplored and detailed investigations are
therefore needed to better clarify the scenario of coexistence of two families
of compact stars, hadronic stars and quark stars, proposed in
\cite{Drago:2013fsa,Drago:2014oja}.

\subsection*{Acknowledgement}

We express our thanks to the organizers of the CSQCD IV conference for providing an 
excellent atmosphere which was the basis for inspiring discussions with all participants.
We have greatly benefited from this.

\end{document}